# Collaborative Problem Solving on a Data Platform Kaggle*

Teruaki Hayashi[1], Takumi Shimizu[2], Yoshiaki Fukami[3]
[1]School of Engineering, The University of Tokyo, Tokyo, Japan
[2]Faculty of Policy Management, Keio University, Japan
[3]School of Medicine, Keio University/ Faculty of Economics, Gakushuin University, Japan
[1]hayashi@sys.t.u-tokyo.ac.jp

*Abstract*—Data exchange across different domains has gained much attention as a way of creating new businesses and improving the value of existing services. Data exchange ecosystem is developed by platform services that facilitate data and knowledge exchange and offer co-creation environments for organizations to promote their problem-solving. In this study, we investigate Kaggle, a data analysis competition platform, and discuss the characteristics of data and the ecosystem that contributes to collaborative problem-solving by analyzing the datasets, users, and their relationships.

## I. Introduction

In recent years, there have been heightened expectations regarding the creation of new businesses and the raising of the value of existing services by exchanging data in various fields. It has reached the point where the conversion into data and circulation of information that had been hoarded inside organizations until now are attracting attention as sources of new innovations. Moreover, it is not always true that data and analysis are completed by a single organization or individual and a specific corporation controls the distribution and exchange of data, as was the case in previous data strategies; instead, the number of services that employ as their foundation a dispersed architecture whereby a variety of stakeholders coordinate flexibly as needed is increasing [1]. Amidst such changes, the deployment of a cooperative creation environment in which data and knowledge are exchanged and problem solving is carried out between different organizations has begun. An ecosystem that employs data as the medium is gradually being formed.

However, even though there is increasing concern about value creation through data circulation (exchange/trade/buy and sell) on platforms, understanding is limited when it comes to coordination and interactions between the data in the data exchange ecosystem. Accordingly, in this study, an attempt was made to understand the structural features possessed by platforms and to clarify the elements and mechanisms that form the data exchange market and ecosystem. If the mechanisms that bring about data exchange and circulation can be derived from the features of a platform, insight can be obtained into the functions, interactions of users, and data use and application that are required in the nascent data exchange market, as well as the methods for deriving value therefrom. This could serve as useful knowledge for the burgeoning data society. In this study, the subject of analysis is the data competition service Kaggle, which is one of the platforms in the data exchange ecosystem. The features of the data and ecosystem that contribute to collaborative problem solving are discussed based on the relationships between data, users, and data coordination and analysis.

In this paper, in Section 2, research on data platforms, which are one form of data ecosystem for promoting data cooperative creation in different types of business, and various studies on collaborative problem solving are surveyed. In Section 3, the approach and experiments of this study are described. In Section 4, the experimental results are discussed, and data cooperative creation and problem solving are examined. Finally, the conclusions are given in Section 5.

## II. Related Studies

### A. Collaborative problem solving

It may well be possible for a solution to be promoted and the value of the data to be determined by data understanding by a single agent. However, the ways in which switching of viewpoints and knowledge acquisition are advanced by the interactions of several agents with a variety of goals and inclinations and how this leads to collaborative problem solving have been widely studied in the field of cognitive science [2–5]. As goes the old saying, "Three heads are better than two." It is believed that discovery of new ways to use the data in question and the promotion of problem solving in a manner different from previous approaches are carried out by communication between the stakeholders of the data exchange market and the users of a platform. In fact, a piece of data that does not necessarily have any meaning or value by itself may well bring about the emergence of meaning and value when it is combined with other data. Kokuryo argues that connecting the data and information on a platform can lead to collaboration between people and the creation of new value [6]. In addition, workshops that promote innovation by spurring communication between the different stakeholders in the data market have been proposed, and data coordination and value creation have been implemented [7, 8].

### B. Data platforms and Kaggle

As noted above, rather than the collection and analysis of, and decision making based on, data by a single organization, data circulation and exchange have begun to be carried out between various entities. The venue with the functions for mediating such data exchanges and transactions is the data platform. The number of corporations offering platform

---

services that support data exchanges has been growing globally, and the participation of business operators whose core business is not data has also been observed in recent years [9]. At the same time that these business operators cooperate and compete with one another, stakeholders are involved in the planning for each service, and an ecosystem centered on the data is gradually being formed owing to the wide range of data exchanges that are being done.

Kaggle is a platform service for data analysis competition that was established in 2010. Corporations and data scientists provide the data and challenges, and users propose such things as analysis models. At times, a monetary award has been arranged in the competition, and this is a system wherein the corporations and institutions that present the challenges purchase outstanding analytical models in exchange for prize money. Moreover, the method for analyzing the datasets that is provided not only by the competition function but also the function of the kernel is made public. In the forum, it is provided with both a function for communication between users about the datasets and kernels and a function for following favorite users. It has become one of the largest data analysis communities in the world. A kernel is an environment that can execute data input, an analysis process, output, etc. on a browser in notebook format, which is the minimum unit of an analysis project in Kaggle. A user can actually execute the code simply on the browser without building an analysis environment with the user's own personal computer. In addition, the kernels can be disclosed to other users, and the source code of the analysis can be shared with the participants in the competition.

Several studies related to the platform for data ecosystem understanding have been conducted, and the characteristics of the data that are handled bit by bit and the mechanisms for the coordination thereof have been clarified [10–12]. However, there are not enough data, users, and platform services with a community to create an ecosystem in the nascent data exchange market, and there have been limits on research to date. In the present study, Kaggle, for which the number of users and the data volume that is handled have been increasing and growing every year as of December 2020, is regarded as one form of the data circulation market. Here, the features of the ecosystems associated with data are examined.

## III. EXPERIMENT

In the experiment, we used the Meta Kaggle dataset[2] (hereinafter, "Kaggle data") provided by Kaggle (downloaded on December 1, 2020). The file names of the datasets and the variable names of the datasets used in the experiments are shown in Table I. The dataset has 65,274 pieces of data, and 1,774,725 kernels that use the datasets, including different versions, are present. Of these, a dataset of 47,619 pieces of data for which version management has been undertaken and that are used in the kernels was taken as the analysis target.

---

[2] https://www.kaggle.com/kaggle/meta-kaggle

TABLE I. VARIABLES OF KAGGLE DATASETS USED IN THE EXPERIMENT

| Data name | Variables |
|---|---|
| Users.csv | Id, UserName, PerformanceTier |
| Datasets.csv | Id, CreatorUserId, CurrentDatasetVersionId |
| DatasetVersions.csv | Id, DatasetId, DatasourceVersionId, CreatorUserId, VersionNumber, Title |
| Kernels.csv | Id, DatasetId, DatasourceVersionId, CreatorUserId, VersionNumber, Title |
| KernelVersionDatasetSources.csv | Id, SourceDatasetVersionId KernelVersionId |
| KernelVersions.csv | Id, AuthorUserId, VersionNumber, Title |

## IV. RESULTS AND DISCUSSION

### A. Features of Users

When the number of registered users of Kaggle was checked, we found that there were 5,832,199 accounts and that these were involved in data provision, planning of competition, discussions about data, analysis, and so on. In other words, it is not necessarily the case that all the users are competition participants. There are also users who have only acquired accounts and users of only data provision. It was also discovered that a division of labor had arisen in a spontaneously generated manner, even though this had not been clearly specified by users.

When the follow–follower relationship between users was investigated, we found that the number of users of the maximum connected component of the network was 170,081, and the secondary distribution thereof was a power distribution (exponent: 2.12; coefficient of determination: 0.994). In other words, the follow–follower network has a structure in which almost no users have any relationship, and a portion of the users have acquired huge links.

Next, we compared the extents of the contributions by users in the platform. In the Kaggle data, the contributions of users are provided by a six-point scale from 0 to 5 as PerformanceTier (this means that, the higher the number, the higher the rank of the user). The extents of the contributions to the platform were compared using this evaluation. Note that the assessment by scores of 0 to 5 does not correspond to Novice, Contributor, Expert, Master, and Grandmaster.

First, it was learned that there were 5,730,037 users with rank 0. This accounted for 98.2% of all users. However, the number of users with ranks of 3 to 5, which indicates the highest level of contribution, was a mere 0.3% of the whole. In other words, the vast majority of Kaggle users had just registered or were lurkers studying the data analysis method, and almost all the activity on the platform involved an extremely small number of users with a high degree of contribution. Such an extreme bias in the user contribution also matches the behavioral pattern that is repeatedly observed in various online communities [13].

Next, we examined the issue of the extent to which the users of each rank in the follow–follower relationship were followers. Figure 1 shows the average number of followers of the users for each rank. Although the number of users with a rank of 0 accounted for 98.2%, the average number of followers was extremely small at 0.0178. However, it was 874.8 for users with a rank of 4, and 228.7 for users with a rank of 5; thus, these users had an extremely large number of followers compared with the other ranks. In other words, the users with a rank of 3 or higher are being followed by many users, and knowledge about data analysis is being exchanged. Interestingly, rather than users with a rank of 5, whose contribution is extremely high, it is instead users with a rank of 4 who have almost four times as many followers. In addition, when we examined the question of which users with which rank were being followed by the users of each rank, the results were as shown in Table II. There was a marked tendency for users to follow users with a rank higher than their own rank. In particular, the number of times that a user of rank 0 followed a user of rank 4 was highest at 107,739, after which came the number of times a user of rank 1 followed a user of rank 4, 64,048. This result suggests that it was mainly users with a rank of 4 that were promoting the interactions between users on the Kaggle platform.

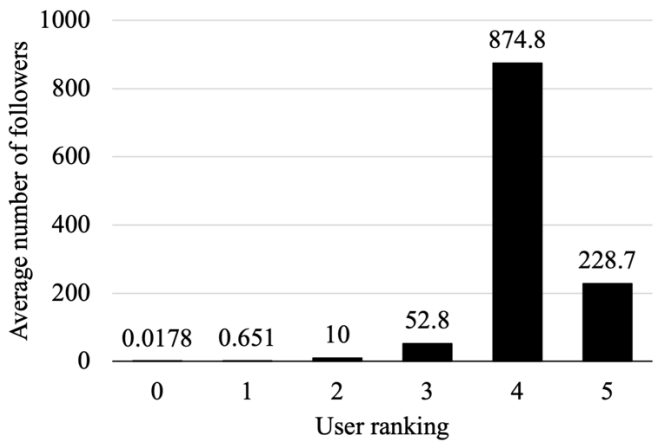

Fig. 1. Average number of followers of users of each rank

TABLE II. NUMBER OF RANKS BEING FOLLOWED BY USERS

|  | All data |
|---|---|
| Ranks higher than one's own rank | 383,306 |
| Ranks that are the same as one's own rank | 100,671 |
| Ranks lower than one's own rank | 39,414 |

## B. Features of datasets and kernels

As the features of datasets and kernels, we found that, of the 65,195 datasets, the number of datasets used in the kernels is 47,619, and 27.0% of the datasets are not used in the kernels even once. Accordingly, when the distribution was examined, it was found that the use frequency distribution of the dataset showed a power distribution (exponent: 2.00; coefficient of determination: 0.977). In other words, it had the characteristic that there were no data that were employed on average for the datasets of the Kaggle platform, and most of them were used just a few times; however, a small number of specific datasets were employed a huge number of times. Phrased differently, while there are "popular data" that are used actively in data analysis, there are also data that are used not even once, even though they are registered, and a large deviation has arisen between datasets.

Figure 2 shows the top-10 datasets in terms of the highest frequency of use in kernels. The top-two items involve COVID-19-related data. In addition, because Iris Species and Wine Reviews are data that are frequently employed by tutorials for data analysis, it can be understood why the use frequency is relatively high. However, what merits close attention here is the fact that the Novel Corona Virus 2019 Dataset, which has the highest use frequency, is relatively new data that were made public from January 30, 2020. In addition, the data registered from January 2020 onward appear in the higher ranks—for example, the COVID-19 Open Research Dataset Challenge (CORD-19) in second place and the COVID-19 Dataset in ninth place. Of all datasets, it was learned that there are 935 that include "COVID-19," "Corona," and "SARS-CoV-2" and that these were being frequently updated and analyzed as of December 2020. In other words, despite the datasets related to COVID-19 being new data, they are used to such an extent that the number of times they have been used by kernels reached the top in a short period. Therefore, even a dataset for which a long time has passed after it was made public can be utilized frequently in a short period provided that it meets the needs of trends and society. The above suggests that the use value of and degree of attention to the data vary greatly depending on the needs.

However, when the use frequency distribution of the datasets before December 31, 2019, before the COVID-19 related data were registered, was examined, it was similarly a power distribution (exponent: 1.91; coefficient of determination: 0.967). This means that, even before the data related to COVID-19 appeared, there was no change in the structure in which a small number of datasets are analyzed an extremely large number of times in data use and application, and most of the datasets are employed only a few times. The above discussion shows that the use frequency distribution of the data in data use and application is usually a power distribution. Even though it hardly changes at all macroscopically, the frequency of the application of individual pieces of data always fluctuates.

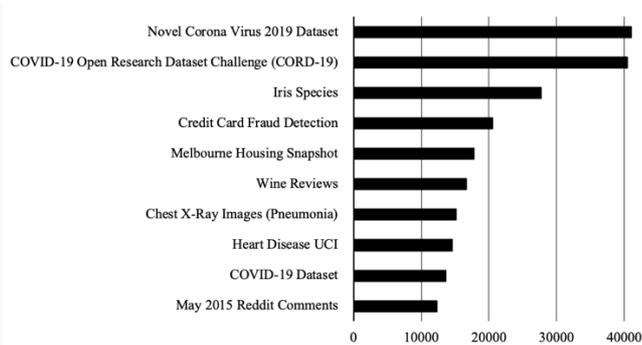

Fig. 2. Top-10 frequently used datasets

## V. CONCLUSION

### A. Summary

In this study, we focused on Kaggle, a data analysis competition service for understanding the ecosystem mediated by data. We discussed the relationship among data, users, and data use and application. The results suggest that, while a portion of the users is attracting the attention of many users, such that the follow–follower relationship of the users is a power distribution, an environment in which pairs of users of different ranks are interacting on the platform may well be making the Kaggle platform function as an ecosystem. In addition, we found that the data that are used from the standpoint of data use and application are a power distribution, there are no exceptional or typical data present in a well-controlled platform, and there is an extreme bias between the data.

### B. Future Work

In this study, we discussed the datasets employed in the kernels, but there are times when multiple datasets are employed in one kernel, and it is known that discovery of knowledge is being achieved by data coordination. However, what is important here is the question of how the data are being coordinated. In fact, even though metadata have been provided in the Kaggle data utilized in this study, the structure thereof has not been fully provided, and it took more time than necessary to create an integrated dataset. When a plurality of data are handled by a kernel, it is necessary to understand what kinds of variable inside the dataset are being connected for what sort of purpose to derive the analysis results.

In addition, aspects of the data that are attracting attention changed greatly with the onset of the global pandemic because of infections by the novel coronavirus at the start of 2020. Data are an economic good that is not consumed, but it is believed to be extremely sensitive to social circumstances and trends. In other words, in future research, it will probably be necessary to give due consideration to the demands of actual society when discussing the mechanisms that result in a data exchange ecosystem having functions as a platform or market, and also the characteristics of the data that are handled thereon. Moreover, a dynamic analysis that considers the time axis will perhaps be necessary when it comes to the phenomenon that users with a rank of 4 have obtained more followers than users with a rank of 5. The reason for this is that static networks rarely exist to begin with in society, which is a macro phenomenon, and these are, to state it in an extreme fashion, approximations of dynamic networks. Analysis that considers the process whereby new data and users are participating at each moment and the dynamics of the community are likely to be crucial in the future.

ACKNOWLEDGMENT

This study was supported by JSPS KAKENHI (JP20H02384).